\documentclass[aps, prd,nofootinbib,superscriptaddress,reprint,onecolumn,notitlepage,11pt,twoside]{revtex4-1}
\usepackage[ includeheadfoot,paperheight = 269mm,paperwidth = 190mm, top=18mm, left = 28mm, right=18mm, bottom=22mm]{geometry} 
\usepackage{graphicx}
\usepackage{amssymb}
\usepackage{amsmath}
\usepackage{fancyhdr}
\usepackage[british]{babel}
\usepackage{amsmath,graphicx,color,mathrsfs,subfigure}
\usepackage{latexsym}
\usepackage{amsmath}
\usepackage{amssymb}
\usepackage{graphics}
\usepackage{epsfig}
\usepackage{graphicx}
\usepackage{subfigure}
\usepackage{bm}

\usepackage{multirow}

\newcounter{contribution}
\numberwithin{equation}{contribution}

\def\be{\begin{equation}}
\def\ee{\end{equation}}
\def\bea{\begin{eqnarray}}
\def\eea{\end{eqnarray}}

\def\mchi{m_{\chi}}

\def\fnl{f_{\rm NL}}
\def\nfnl{n_{f_{\rm NL}}}
\def\gnl{g_{\rm NL}}

\def\taunl{\tau_{\rm NL}}

\def\fnlloc{f_{\rm NL}^{\rm local}}
\def\fnleq{f_{\rm NL}^{\rm equil}}
\def\fnlorth{f_{\rm NL}^{\rm ortho}}
\def\la{\langle}
\def\ra{\rangle}

\def\bkone{\mathbf k_1}
\def\bktwo{\mathbf k_2}
\def\bkthree{\mathbf k_3}

\def\picube{(2\pi)^3}
\def\bk{{\bf k}}
\def\bkp{{\bf k'}}

\def\kpivot{k_{\rm pivot}}

\newcommand{\sdelta}[1]{\!\delta^{\,3}(\mathbf{#1})}
\def\Mp{M_{\rm Pl}}

\def\rdec{r_{\rm dec}}

\begin{document}
\setcounter{page}{1}


\title{Lecture notes on non-Gaussianity}

\author{Christian T.~Byrnes\footnote{E-mail: c.byrnes@sussex.ac.uk}}
\affiliation{Department of Physics and Astronomy, Pevensey II Building, University of Sussex, BN1 9RH, UK}

\begin{abstract}

We discuss how primordial non-Gaussianity of the curvature perturbation helps to constrain models of the early universe. Observations are consistent with Gaussian initial conditions, compatible with the predictions of the simplest models of inflation. Deviations are constrained to be at the sub percent level, constraining alternative models such as those with multiple fields, non-canonical kinetic terms or breaking the slow-roll conditions. We introduce some of the most important models of inflation which generate non-Gaussian perturbations and provide practical tools on how to calculate the three-point correlation function for a popular class of non-Gaussian models. The current state of the field is summarised and an outlook is given. 

\end{abstract}

\maketitle


\tableofcontents

\section{Introduction and the aims of these lecture notes}

The theory of inflation, a period of quasi-exponential expansion of the universe very shortly after the big bang is now widely regard as part of the standard cosmological model. The predictions of the simplest inflationary models have passed increasingly stringent tests from observations of the cosmic microwave background (CMB), most recently by the Planck satellite. Remarkably, the apparently crazy idea that the formation of all structures in our universe such as galaxies were caused by quantum perturbations of the field driving inflation, does have strong observational evidence. Inflation therefore provides a mechanism to relate the smallest and largest scales in the universe.

Despite the evidence that inflation occurred, rather little is known about the properties of inflation. The energy scale could be anywhere between the TeV and the GUT scale, an enormous range, notable for stretching far beyond the highest energies we can ever reach with a terrestrial experiment such as a particle collider. This provides cosmologists with the opportunity to provide constraints on extremely high energy physics, as witnessed by the study on the field of embedding inflation into models of string theory. It is both an opportunity and a challenge that we have limited information about the relevant model of particle physics during inflation. For example, was inflation driven by one or more fields, what form did their potential(s) take, and their kinetic term(s)? What is the energy scale of inflation, and how did the universe become radiation dominated after inflation ended? 

Our best way to answer these questions is by studying the statistical properties of the perturbations generated during inflation for different classes of models and then evolving this spectrum of perturbations forwards in time to make predictions for the pattern of temperature perturbations in the CMB, as well as the density perturbations of large scale structure such as galaxy clusters. 

Over the past decade, the study of the Gaussianity of the primordial perturbations has become a large field, being the main theme of many focused conferences and workshops every year \cite{Komatsu:2009kd}. The simplest models of inflation are expected to produce perturbations which are extremely close to Gaussian \cite{Maldacena:2002vr}. Any observation of non-Gaussianity would rule out the simplest models. Gaussian perturbations are very tightly constrained by the definition of Gaussianity, for example all information about the correlation functions of the perturbations is encoded in the two-point function alone. By contrast, non-Gaussian perturbations could be anything else. This  opens a Pandoras box full of possibilities to search a huge observational data set for anything and everything, which leads to the danger of coincidental patterns in the data being interpreted as signals of real physics. Anomalies of this sort, strange patterns which were not predicted in advance of analysing the data do exist, but all are of a reasonably small statistical significance, especially when taking into account the ``look elsewhere'' effect, i.e.~if 100 people search for different non-Gaussian patterns in a Gaussian data set, then one of these patterns will probably be ``detected'' at the $99\%$ confidence level due to the statistical fluctuations in the data.

Fortunately general classes of inflationary models do predict specific shapes of non-Gaussianity, for example certain templates of the bispectrum  (the three-point function, which is zero for a Gaussian distribution) would point towards a requirement that multiple fields contributed to the physics of inflation, while a non-canonical kinetic term of the inflaton fields Lagrangian predict a different signature in the bispectrum which is observationally distinct. Classifying different models according to the form of non-Gaussianity they generate, and finding observational constraints on the corresponding scenario has become a major topic within the study of inflation.

The aim of these lectures is to provide both a background knowledge about non-Gaussianity of the primordial perturbations which gave rise to all structures as well as some concrete calculations for reasonably simple scenarios, thereby developing a real working knowledge of the field and providing the tools to perform calculations yourself for some scenarios.  The course contents are up to date and much of the content here is not contained in any textbook, at least at the time of publishing these notes.


These lecture proceedings are based on a four hour lecture course held at the II JBP Cosmology school in Espirito Santo, Brasil. The slides are available from the school's website (http://www.cosmo-ufes.org/jpbcosmo2-mini-courses-seminars--posters.html). The most closely related courses were on Inflationary Cosmology by J{\' e}r{\^ o}me Martin and CMB theory by David Wands. 

Students wishing to learn more may turn to many review articles about non-Gaussianity, here we just list a selection published during or after 2010. However note that all predate the Planck data release which significantly improved the constraints on non-Gaussianity. Reviews which are more focused towards how the observational constraints are made include \cite{Komatsu:2010hc,Yadav:2010fz,Liguori:2010hx,Desjacques:2010nn,Verde:2010wp,Bartolo:2010qu} and focus on tests relating to isotropy and anomalies  are considered in \cite{Abramo:2010gk,Copi:2010na}.   Perhaps the most comprehensive review of non-Gaussian inflationary models is given in \cite{Chen:2010xka}. Reviews focused on the local model of non-Gaussianity include \cite{Wands:2010af,Byrnes:2010em}

The plan of the lectures is as follows: We first provide more introduction and motivation for this topic, presented in a novel manner. We then study some specific models of non-Gaussianity, and the classes of models which give rise to them, Sec.~\ref{sec:models}. In section \ref{sec:deltaN}  we provide a practical introduction to a simple method of calculation inflationary perturbations to non-linear order, the $\delta N$ formalism and use this to derive some general formulae which are useful for calculating the amplitude of the three and four-point functions of the inflationary perturbations. We then go into great detail to study a concrete example of a popular inflationary model whose perturbations are non-Gaussian, the curvaton scenario, see Sec.~\ref{sec:curvaton}. In section \ref{sec:faqs} we provide some nontechnical question and answers about the subject and its current status after the first major data release from the Planck satellite, which has made by far the most stringent constraints on non-Gaussianity available. Finally we conclude in Section \ref{sec:conclusions}.

\section{Gaussian distributions}

A Gaussian (or normal) distribution is defined by the probability distribution function (pdf)
\bea \frac{1}{\sqrt{2\pi\sigma}} e^{-\frac{\left(x-x_0\right)^2}{2\sigma^2}}, \eea
where $x_0$ is the mean of the distribution and $\sigma^2$ the variance. Gaussian distributions are relatively simple and have many neat properties, some of which we will explore in these lectures. It only has two free parameters, and we will see that in cosmology the mean can usually be redefined to be zero, leaving only the variance. In contrast, non-Gaussian distributions can have an arbitrary number of free parameters.

The central limit theorem states that in the limit of a large number of measurements, if all measurements are drawn from independent, identically distributed pdfs then the limiting distribution will the Gaussian. Because many processes in nature depend only on the average of many ``small scale" processes, we often find nearly Gaussian perturbations in nature. Therefore a good first guess for an unknown distribution is often that it will be nearly Gaussian. 

In quantum mechanics, the ground state of the simple harmonic oscillator is Gaussian distributed. Since inflation predicts that the primordial density perturbation is generated from the quantum fluctuations of a light scalar field, which is quantised analogously to the simple harmonic oscillator, we may expect the curvature perturbation to be Gaussian. 

This means, for example, that if you divide the CMB sky into many small patches, there will be the same number of patches which are hotter than average as those which are colder than average, and that if you plotted a histogram of the temperature deviations in each patch, they would form a normal (bell) curve. Similarly if you were to divide the early universe into little cubes, the density distribution would fit a Gaussian curve. However, doing the same in the late time universe, when the density perturbations have become large this changes, since
\bea -1\leq \frac{\rho-\bar{\rho}}{\bar{\rho}}\leq \infty, \eea
where $\bar{\rho}$ is the average density over the total volume, and $\rho$ the density in the patch being measured,
is not a symmetric distribution, unlike the Gaussian distribution. The asymmetry arises because there is a limit to how empty space can become ($\rho=0$), whilst there is almost no limit to how overdense it can become. 

There are two lessons which can be learnt from this. Firstly that small perturbations can more easily be Gaussian in practice, for example the CMB temperature is given by $T=2.75\pm10^{-5}$K, and so although the temperature cannot be lower than absolute zero, in practice no perturbation will ever come close to being so large. We will later see that the temperature distribution of the CMB comes extremely close to following a Gaussian distribution. Secondly gravity acts both to make the density perturbation larger (overdense regions become denser due to gravitational attraction, leaving the voids emptier) and less Gaussian. A Gaussian distribution remains Gaussian under a linear transformation (which corresponds to shifting the mean value or changing the variance), but becomes non-Gaussian under any non-linear transformation, for example squaring a Gaussian distribution leads to a chi-squared distribution. In practise, we will always identify the linear perturbation with a Gaussian perturbation, and all higher-order terms as being non-Gaussian corrections. The lowest order corrections, the quadratic corrections, follow a chi-squared distribution.

Since the gravitational equations of motion are non-linear (thats what makes them hard to solve!), the perturbations will not only grow, but they will also become non-Gaussian, even if the initial perturbations were exactly Gaussian. The non-Gaussianity generated if one starts with a Gaussian primordial density perturbation is known as the secondary non-Gaussianity. Until the time that the CMB formed, this secondary non-Gaussianity was very small and the corrections can be neglected. For large scale structure, the secondary non-Gaussianity becomes large at later times, and at smaller scales which have a larger amplitude and are hence less linear. This means that detecting primordial non-Gaussianity is easier in the CMB than large scale structure; most (or all) of the non-Gaussian signal measured in the clustering of galaxies is secondary non-Gaussianity. These lectures will focus exclusively on primordial non-Gaussianity, which we aim to use as a probe into the physics of the early universe. Fundamental questions which we hope to answer include how many scalar fields were present during inflation, what form their Lagrangian had and how reheating proceeded after inflation.

\subsection{Distinct characteristics of Gaussian distributions}

As described above, a Gaussian distribution has just two free parameters, the mean and variance. The mean can typically be defined to be zero, since it is convenient to define a perturbed quantity as being its deviation from the average value. Physically the mean tells us about the homogeneous universe, but nothing about the primordial perturbations. For example, the mean density of the universe has been measured as being very close to the critical density, which corresponds to a spatially flat universe. The average value of the CMB temperature redshifts with time and hence provides us with no information about inflation. The only additional information we can learn for a Gaussian distribution is how the variance depends on scale, for example whether the variance becomes larger if we divide the CMB sky into larger patches. In practise such a measurement is usually discussed in Fourier space, where the Fourier wavenumbers satisfy $k=|{\bf k}|\sim({\rm physical\;scale})^{-1}$, and the two-point correlator of the curvature perturbation is related to the power spectrum by 
\begin{equation}\label{2ptfn}
\la \zeta_\bk \zeta_\bkp \ra = P_\zeta(k)\picube \sdelta{\bk+\bkp} \,,
\end{equation}
and the variance per logarithmic interval in $k$-space is given by
\begin{equation}\label{P}
{\cal P}_\zeta(k) = \frac{4\pi k^3}{(2\pi)^3} P_\zeta(k)=A_s \left(\frac{k}{\kpivot}\right)^{n_s-1}.
\end{equation}
$A_s$ denotes the amplitude of the scalar perturbations, it corresponds to the variance of the perturbations at the pivot scale. The spectral index, $n_s-1$ parametrises a possible scale dependence, where $n_s=1$ corresponds to scale independence (in which case the power spectrum does not depend on $k$). 
Using combined Planck and WMAP data, see table 2 of \cite{Ade:2013zuv}, the values of the two primordial spectra parameters are
\bea \ln\left(10^{10}A_s\right)  = 3.089^{+0.024}_{-0.027},   \\ n_s-1= 0.9603\pm0.0073.  
\eea

The assumptions of homogeneity and isotropy imply that the power spectrum is only a function of $k$. The parametrisation (\ref{P}) is a simple ansatz for the scaling which has got nothing to do with whether the perturbations follow a Gaussian distribution or not. This simple ansatz is a good match to observations, meaning that we only require two parameters to describe the primordial power spectrum. If the perturbations are Gaussian, then all of the information is contained in the two-point correlation function, or equivalently the power spectrum. All of the odd n-point correlators are zero, while all of the even n-point correlators can be reduced to disconnected products of the two-point function, which contain no new information (this is known as Wicks theorem, we will not prove it here, it is a standard proof in quantum field theory courses). Hence Gaussian statistics are very prescriptive but not very informative, everything can be learnt just by measuring the two-point function. Given that we have not detected primordial non-Gaussianity, and we can parametrise the power spectrum using just two numbers, we face the remarkable fact that observations of millions of pixels on the CMB sky, which lead to over one thousand well measured power spectrum amplitudes (the $C_l$'s), can be described by primordial perturbations which are fully specified by only two parameters. Such a simple state of affairs is perfectly consistent with inflation, a period of quasi de Sitter expansion being driven by a single slowly-rolling scalar field can naturally lead to a spectrum of nearly Gaussian and nearly scale invariant perturbations.

In order to learn more about inflation, we must carefully check how consistent this simple picture is, both theoretically and as a match with observations. The rest of these lectures will hopefully motivate why the difficult search for non-Gaussianity is worthwhile and exciting, despite the lack of any clear observational detection of primordial non-Gaussianity.

\section{Different models of non-Gaussianity}\label{sec:models}

As described in the last section, Gaussianity is very prescriptive. Non-Gaussianity is anything else, so in principle we should search for all possible patterns in the data when hunting for non-Gaussianity. Apart from the fact that this is computationally unfeasible, it is also clear that there will be random patterns distributed in any large data set, many of which are statistical flukes of no cosmological significance. If the underlying distribution is Gaussian, then we should still expect one in a hundred tests to appear non-Gaussian at the $99\%$ confidence level. This is the well known issue about anomalies in the CMB  and the difficulty of quantifying the unlikeliness of posterior distributions. For example see the review article about large angle anomalies \cite{Copi:2010na}  and the WMAP paper \cite{Bennett:2010jb}, which often do not reach the same conclusions. 

The path we will take in these lectures is to study which types of non-Gaussianity simple models of inflation predict. A major advance from the last decade in this field is the realisation that different types of extensions of the simplest inflationary models produce specific and predictable types of non-Gaussianity \cite{Komatsu:2009kd}. Models of single-field, slow-roll inflation with canonical kinetic terms and a Bunch-Davies initial vacuum state produce Gaussian perturbations. Breaking any one of these four conditions produce specific shapes of non-Gaussianity. These shapes are then searched for with data and the constraints can be interpreted in terms of model parameters, for example constraining how strongly the kinetic term of the inflaton field is allowed to deviate from a canonical form. 

In analogy with the two-point function which defines the power spectrum, (\ref{2ptfn}), we define the bispectrum via the three-point function of the curvature perturbation
\bea\label{3ptfn} \la\zeta(\bkone)\zeta(\bktwo)\zeta(\bkthree)\ra = (2\pi)^3\sdelta{\bkone+\bktwo+\bkthree}B_\zeta(k_1,k_2,k_3). \eea
The delta function comes from assuming statistical homogeneity, assuming isotropy in addition allows us to write the bispectrum in terms of just the three amplitudes of the wave vectors (i.e. the three side lengths of a triangle in Fourier space). Compared to the power spectrum which was a function of just one amplitude, we see that the bispectrum may contain a lot more information, i.e.~information about its shape as well as about its amplitude. 

\subsection{Local non-Gaussianity}

We have previously described how the linear perturbations can be identified as the Gaussian perturbations. An obvious next step would be to consider the linear term as the first in a Taylor series expansion, in which the second order term is of order the linear term squared, and so on for higher orders. Given that the amplitude of  the linear term is observed to be $10^{-5}$, we may expect the convergence to be strong, and that we are unlikely to need to include terms up to a high order. This approach is not only mathematically quite simple, but also quite well motivated by many classes of physical models. We first study how the model is defined, and we provide a technique for calculations before providing examples of concrete models. In particular, we will make an in depth study of the curvaton scenario, and use this case both as a justification of the local model, and to motivate extensions to this simple model. We will then see how non-Gaussianity provides an observational probe to distinguish between different models of inflation.

The local model of non-Gaussianity, in its simplest form, is defined by
\bea
\label{zetaloc}
\zeta({\bf x})=\zeta_G({\bf x}) + \frac35 \fnl( \zeta_G^2({\bf x})-\langle\zeta_G^2({\bf x})\rangle ).
\eea
The name comes from the fact that it is defined locally in real space, $\zeta$ is a local function of position. The factor of $3/5$ comes fom the original definition being in terms of the Bardeen potential \cite{Komatsu:2001rj}, which is related on large scales to the primordial curvature perturbation in the matter dominated era by $\Phi=(3/5)\zeta$. The variance term has been subtracted from the quadratic part in order that the expectation value satisfies $\la\zeta\ra=0$, any other choice would leave a non-zero expectation value, which would be degenerate with the background term, meaning that $\zeta$ would not be a purely perturbed quantity. 
When working in Fourier space, this constant term is only important for the $k=0$ mode (zero wavelength corresponds to a homogeneous mode), and is therefore often neglected. Under a Fourier transform, the quadratic term becomes a convolution, 
\bea
\label{zetalocF}
\zeta({\bf k})=\zeta_G({\bf k}) + \frac35 \fnl \frac{1}{\picube} \int d^3 {\bf q} \zeta_G({\bf q}) \zeta_G({\bf k}-{\bf q}).
\eea
 
In general, the local bispectrum is defined by
\bea B_\zeta^{\rm local}=2 \frac35\fnlloc\left( P_\zeta(k_1)P_\zeta(k_2) +2\;{\rm perms}\right),\label{Bloc} \eea
which in the case of a scale-invariant power spectra reduces to
\bea B_\zeta^{\rm local}=2A_s^2\fnlloc\left\{\frac{1}{k_1^3 k_2^3}+2{\rm\; perms}\right\}.  \eea
Although (\ref{zetaloc}) does imply (\ref{Bloc}) the reverse is not the case, more general models for $\zeta$ can give rise to the same bispectrum. We will study generalisations of the local model in Sec.~\ref{sec:local-extensions}, and we will see how i) the scale dependence of $\fnl$ in Sec.~\ref{sec:scale-dep} and ii) the trispectrum in Sec.~\ref{sec:trispectrum} may be used to break the degeneracy between these different expansions for $\zeta$ which generate equal amplitudes (and shapes) of the bispectrum.

\subsection{Equilateral and orthogonal shapes}

If the inflaton field has a non-canonical kinetic term, interaction terms may give rise to a large bispectrum. The bispectrum is maximised for three modes which have the strongest interaction, for this model this corresponds to the time when  all three of the modes exit the horizon during inflation, and hence all have similar wavelengths. Two templates which are both maximised in the limit of an equilateral triangle have been commonly used as a test of such models are the equilateral and orthogonal, the latter was designed to be orthogonal to the equilateral \cite{Senatore:2009gt}, as the name suggests.
\begin{align} B_\zeta^{\rm equil}=6A_s^2\fnleq\left\{-\left(\frac{1}{k_1^3 k_2^3}+2\;{\rm  perms}\right)-\frac{2}{(k_1 k_2 k_3)^2}+\left(\frac{1}{k_1k_2^2k_3^3}+5\;{\rm perms}\right)\right\},  \\ 
B_\zeta^{\rm ortho} = 6A_s^2\fnlorth \left\{-3\left(\frac{1}{k_1^3 k_2^3}+2\;{\rm perms}\right)-\frac{8}{(k_1 k_2 k_3)^2}+3\left(\frac{1}{k_1k_2^2k_3^3}+5\;{\rm perms}\right)\right\}.
\end{align}
For both cases we ignore the scale dependence of the power spectrum and the intrinsic scale dependence of the bispectrum itself \cite{Chen:2005fe}. 

Popular models of inflation with non-canonical kinetic terms include k-inflation \cite{ArmendarizPicon:1999rj} and Dirac-Born-Infeld (DBI) inflation \cite{Silverstein:2003hf,Alishahiha:2004eh}. DBI inflation is one of the most popular string theory inspired models of inflation, for a recent review of the models see \cite{Baumann:2014nda}, and it generically  predicts such a large equilateral non-Gaussianity that some of the ''simplest" realisations have already been ruled out, see e.g.~\cite{Lidsey:2007gq}.

Single field models can be parametrised as ${\cal L}=P(X,\phi)$, where the kinetic term is $X=g^{\mu\nu}\partial_{\mu}\phi\partial_{\nu}\phi$, and a model with canonical kinetic term satisifies ${\cal L}=-X/2-V(\phi)$, and hence has a speed of sound equal to unity, where
\bea c_s^2=\frac{P_{,X}}{P_{,X}+2XP_{,XX}}. \label{cs}\eea
We note that despite the widespread use of the subscript $s$ in the literature, (\ref{cs}) really defines the phase speed of the perturbations, which is the same as the adiabatic sound speed for classical fluids but not for scalar fields. See \cite{Christopherson:2008ry} for a clarification of this point. For models with a sound speed much less than one, one has $\fnl\sim1/c_s^2$ for both the equilateral and orthogonal models, and hence the Planck constraints provide a lower bound on this parameter $c_s^2\gtrsim0.1$ \cite{Ade:2013ydc}. 

\subsection{Feature models}

It is possible to generate large non-Gaussianity in single-field models with canonical kinetic terms if slow-roll is violated. By definition, inflation requires that $\epsilon<1$, but higher-order derivatives of this parameter can become large. They can only become large for short periods of time, otherwise typically $\epsilon$ will quickly grow to become order unity and end inflation before $60$ efoldings of inflation have been achieved. 

A temporary break down of slow-roll can be achieved by adding a step like feature into the potential, for example of the form \cite{Chen:2006xjb}
\bea V(\phi)=V(\phi)_{\rm sr}\left(1+ c \tanh\left(\frac{\phi-\phi_s}{d}\right)\right), \eea
where $V(\phi)_{\rm sr}$ is the potential which generates slow-roll inflation, $c$ is the height of the step, $d$ the width of the step and $\phi_s$ determines the position of the step. Only the modes which exit the horizon while the inflaton is traversing the step will have an enhanced non-Gaussianity, hence the bispectrum will have a localised shape in Fourier space around a characteristic scale, determined by the Hubble scale when $\phi=\phi_s$. If this scale does not fit inside the range of about seven efoldings which the CMB probes there is almost no hope of a detection. The bispectrum typically also has fast oscillations imprinted on it, which makes it observationally hard to detect or constrain \cite{Chen:2006xjb}. However a violation of slow roll will generically also produce a feature in the power spectrum at the same scale, so correlating this feature with the bispectrum should aid a detection in such models. 

\subsection{Other bispectral shapes}

Although the four shape templates already described above are among the most popular, plenty of others exist and have been searched for. Other well known examples include
\begin{enumerate}
\item {\bf Flattened/folded} configuration, which can be generated by models where the initial state of the perturbations is not the usual Bunch-Davies vacuum state, but an excited state \cite{Chen:2006nt,Holman:2007na,Meerburg:2009ys}. The name comes because this shape has the largest signal in the limit of a flattened isosceles triangle, satisfying $k_1\simeq k_2\simeq k_3/2$. 

\item {\bf Cosmic strings} or other topological defects are strongly non-Gaussian objects which generate a complicated non-Gaussian shape, which has still not been fully calculated. For a review see \cite{Ringeval:2010ca}. The Planck constraints on topological defects coming from both the power spectrum and bispectrum are given in \cite{Ade:2013xla}.

\item {\bf Magnetic fields} are ubiquitous in the universe and have been observed to exist in galaxies, galaxy clusters and even in voids. Their origin remains a mystery. Magnetic fields are intrinsically non-Gaussian objects and their abundance is constrained by the observed Gaussianity of the CMB perturbations. Interestingly, the trispectrum is claimed to provide a tighter constraint on the magnetic field abundance than the bispectrum \cite{2012PhRvL.108w1301T}. For a review of this topic see \cite{Durrer:2013pga}.

\end{enumerate}

\subsection{How similar are the bispectral shapes?}

Given that a huge number of bispectral shapes can be generated, many of which are very similar to one of the standard shapes considered earlier in this section, it is useful to be able to calculate how correlated two different shapes are. Since these shapes are a function of three variables, this is hard to do analytically or ``by eye". For example, DBI inflation predicts a bispectral shape which is highly correlated to the equilateral shape, but it is numerically much harder work with since it is not separable in to a product of the three side lengths. Knowing that they have nearly the same shape, one may use the observation constraint on equilateral non-Gaussianity to constrain the sound speed of DBI inflation.

A scale invariant non-linearity parameter corresponds to a bispectrum which scales as $B\propto P^2\propto k^{-6}$, so it is helpful to define a shape function which factors out this momentum dependence
\bea S(k_1,k_2,k_3)= \frac{1}{\fnl} \left(k_1k_2k_3\right)^2 B_\zeta(k_1,k_2,k_3). \eea

The shape correlator is defined as the inner product of two shapes, to which a volume weighting $1/ \Sigma  k_i$ is applied, designed in order to match the signal to noise which comes from experiments, for details see \cite{Fergusson:2008ra}. The inner product between the two shapes $S$ and $S'$ is given by
\bea F(S,S')=\int_{V_k} S(k_1,k_2,k_3)S'(k_1,k_2,k_3)\frac{1}{k_1+k_2+k_3}d V_k, \label{F:shape}\eea
where $V_k$ is the allowed volume in Fourier space of the $k$ modes, which must satisfy the delta function condition $\Sigma {\bf k}_i=0$. It should also satisfy constraints on the largest and smallest modes available to the experiment, defined respectively by the volume and resolution of the survey. In practise, the results are often reasonably insensitive to these choices, and if (\ref{F:shape}) converges when integrated over all $k_i$, this value is sometimes used without applying any cut offs. The shape correlator is defined by
\bea {\mathcal C}(S,S')=\frac{F(S,S')}{\sqrt{F(S,S)F(S',S')}}. \label{C:shape} \eea
The equilateral shape is about $50\%$ correlated with the local shape \cite{Fergusson:2008ra} and uncorrelated with the orthogonal shape \cite{Senatore:2009gt}.

\section{Local non-Gaussianity and its extensions}\label{sec:local-extensions}

We will first provide a convenient method to calculate the curvature perturbation $\zeta$, which is especially convenient for the local model of non-Gaussianity. We will also see how this can be used to study generalisations of the local model to include scale dependence of $\fnl$ in Sec.~\ref{sec:scale-dep}, and to higher order in perturbation theory to study the trispectrum in Sec.~\ref{sec:trispectrum}.

\subsection{The $\delta N$ formalism}\label{sec:deltaN}

The flat, unperturbed FRW metric is given by
\bea ds^2=-dt^2+a(t)^2\delta_{ij}dx^i dx^j , \eea
and neglecting vector and tensor perturbations, the perturbed space-space part is given by
\bea g_{ij}=a(t)^2 e^{2\zeta(t,{\bf x})} \delta_{ij}.  \eea
Therefore, the curvature perturbation $\zeta$ is the difference between the local expansion rate and the global expansion rate
\bea \zeta(t,{\bf x})=\delta N=N(t,{\bf x})-N(t),  \eea
where
\bea  N(t)=\ln\left(\frac{a(t)}{a_{\rm initial}}\right)=\int \frac{da}{a}=\int H(t) dt. \eea
$N$ should be integrated from a spatially flat hypersurface shortly after horizon crossing, to a final uniform energy density  (or equivalently a uniform Hubble) hypersurface. For some references which developed the $\delta N$ formalism to linear order see \cite{Starobinsky:1986fxa,Sasaki:1995aw,Sasaki:1998ug,Lyth:2004gb}.

During inflation, the scalar fields provide the only contribution to the energy density, and within the slow-roll approximation their time derivatives do not provide a second degree of freedom (i.e.~the field value and its derivative are not independent). Therefore
\bea \zeta=N(\phi_a+\delta\phi_a)-N(\phi_a), \eea
where $a$ labels the scalar fields, and we may expand this as a Taylor series to find the key result
\bea\label{deltaN} \zeta= N_a \delta\phi_a+\frac12 N_{ab}\delta\phi_a\delta\phi_b+\cdots, \eea
where the field perturbations should be evaluated at the initial time (shortly after horizon crossing), summation convention is used and 
\bea N_a=\frac{\partial N}{\partial\phi_{a*}}. \eea
Notice that the derivatives of $N$ depend only on background quantities, so provided that the statistical distribution of the field perturbations is known at horizon crossing, we can do perturbation theory using only background quantities. For some intuition about this remarkable fact, see for example \cite{Wands:2000dp}.

Assuming canonical kinetic terms, Bunch Davies vacuum and slow roll, the initial conditions are very simple. The field perturbations are Gaussian, and 
\bea\label{deltaphi-ab} \langle \delta\phi_a(\bk) \delta\phi_b(\bkp) \rangle = \delta_{ab} P_*(k)\picube \sdelta{\bk+\bkp}, \eea
where
\begin{equation}
{\cal P}_*(k) = \frac{4\pi k^3}{(2\pi)^3} P_*(k)=\left(\frac{H_*}{2\pi}\right)^2\,.
\end{equation}
The cross correlation terms in (\ref{deltaphi-ab}) are slow-roll suppressed and hence neglected \cite{Byrnes:2006fr}. Using these results and the linear part of (\ref{deltaN}), we may calculate the power spectrum
\bea\label{P:deltaN} P_\zeta(k)=N_a N_aP_*(k)\,. \eea

To calculate $\fnl$, we first need the three-point function of $\zeta$. The first non-zero contribution comes from taking two first-order terms and one second-order term from (\ref{deltaN}), we arbitrarily take the second-order term to correspond to the $\bkthree$ term, the other two choices correspond to the two permutations which are added
\begin{align} \la\zeta(\bkone)\zeta(\bktwo)\zeta(\bkthree)\ra&=\frac12 N_a N_b N_{cd} \la \delta\phi_a(\bkone)\delta\phi_b(\bktwo) \int\frac{d^3 {\bf q}}{(2\pi)^3}\delta\phi_c({\mathbf q})\delta\phi_d(\bkthree-{\mathbf q})\ra+2\;{\rm perms}  \nonumber \\  &= \frac12 N_a N_b N_{cd} 2 \int\frac{d^3 {\bf q}}{(2\pi)^3}  \la \delta\phi_a(\bkone)\delta\phi_c({\mathbf q})\ra \la  \delta\phi_b(\bktwo) \delta\phi_d(\bkthree-{\mathbf q})\ra+2\;{\rm perms}  \nonumber \\  &= N_aN_bN_{cd}\int d^3{\mathbf q}\delta_{ac}P_*(k_1)\sdelta{\bkone+{\mathbf q}}\la\delta\phi_b(\bktwo)\delta\phi_d(\bkthree-{\mathbf q}) \ra+2\;{\rm perms}  \nonumber  \\ &=N_aN_bN_{cd}\delta_{ac}\delta_{bd}P_*(k_1)P_*(k_2)\sdelta{\bkone+\bktwo+\bkthree})+2\;{\rm perms}  \nonumber \\ & = N_aN_bN_{ab}\frac{1}{\left(N_cN_c\right)^2}P_\zeta(\bkone)P_\zeta(\bktwo) +2\;{\rm perms}.   \end{align}
In going from the first to the second line, we have applied Wick's theorem to split the four-point function into two two-point functions. Placing the $\delta\phi_c$ and $\delta\phi_d$ in to the same angle bracket results in zero unless $k_3=0$, which is not observationally relevant. This leaves us with two ways to get a non-zero result, and both of those cases lead to exactly the same result which explains the factor of 2. We have applied (\ref{deltaphi-ab}) to go to the third line and performed the integration to reach the fourth line. The final line follows by application of (\ref{P:deltaN}). 

Then using the definition of the bispectrum (\ref{3ptfn}), as well as that of $\fnlloc$, (\ref{Bloc}), we find
\bea\label{fnl}  \fnl = \frac{5}{6} \frac{N_a N_b N_{ab}}{\left(N_c N_c\right)^2} \,. \eea
This result was first derived by Lyth and Rodriguez in 2005 \cite{Lyth:2005fi}, and is very useful since it allows us to calculate the bispectrum amplitude using only background quantities (and we know it must have the local shape). 

\subsubsection{Single-field inflation}

In the case of single-field inflation, the derivatives of $N$ are given by
\bea N'&\simeq& \frac{\bar{H}}{\dot{\bar{\varphi}}}
\simeq \frac{1}{\sqrt{2}}\frac{1}{\Mp}\frac{1}{\sqrt{\epsilon}}\sim\mathcal{O}\left(\epsilon^{-\frac12}\right)\,,  \\ 
N''&\simeq&-\frac12\frac{1}{\Mp^2}\frac{1}{\epsilon}(\eta-2\epsilon)\sim\mathcal{O}\left(1\right)\,, \eea
where the slow-roll parameters are defined by
\bea \epsilon=\frac{\Mp^2}{2}\left(\frac{V'}{V}\right)^2 \,,\qquad 
\eta=\Mp^2\frac{V''}{V}\,. \eea
This suggests that
\bea f_{NL}=\frac{5}{6}\,\frac{N''}{N'{}^2}=\frac56(\eta-2\epsilon) \eea
but since $\fnl$ is slow-roll suppressed for this model, we should have also included the equally small non-Gaussianity of the field perturbations at horizon exit. 

The final result, known as the Maldacena consistency relation \cite{Maldacena:2002vr}, states that
\bea\label{Maldacena}
 {f}_{{\rm NL}} &\equiv&\frac{5}{12}\,\lim_{k_1\to0}\frac{B(k_1,k_2,k_3)}{P(k_1)
  P(k_2)}=\frac{5}{12}(1-n_s). \eea
See Creminelli and Zaldarriaga \cite{Creminelli:2004yq} for a general proof, valid for any single field model (even with non-canonical kinetic terms, breaking slow roll and a non Bunch-Davies vacuum state). The exciting result is that a detection of the bispectrum in the squeezed limit (similar to local non-Gaussianity) would rule out all single-field models. A detection of of non-Gaussianity in any non-squeezed configuration would not do this.

\subsubsection{Single-source inflation}

If we instead assume that a single-field generated the primordial curvature perturbation, which was not the inflaton field, then large local non-Gaussianity is possible (but not required or even generic). Many models in the literature fit into this case, for example
\begin{itemize}
\item the curvaton scenario (to be studied in depth in Sec.~\ref{sec:curvaton})
\item modulated (p)reheating (the duration of reheating varies with position) \cite{Dvali:2003em,Zaldarriaga:2003my,Suyama:2007bg,Ichikawa:2008ne}
\item inhomogeneous end of inflation (the duration of inflation varies with position) \cite{Bernardeau:2002jy,Bernardeau:2002jf,Lyth:2005qk,Huang:2009vk}
\item multiple-field slow-roll inflation can produce large non-Gaussianity for certain trajectories which turn in field space \cite{Alabidi:2006hg,Byrnes:2008wi,Peterson:2010mv,Wang:2010si,Elliston:2011dr}
\end{itemize}

What they all have in common is that the duration of periods with differing equations of state varies with position. This is required in order that $N$ becomes perturbed, since it only depends on the amount of expansion, i.e.~$H$. In modulated reheating, the equation of state is $0$ while the inflaton oscillates in a quadratic potential, but jumps to $1/3$ after the inflaton has decayed into radiation. This means that varying the time of reheating will change the expansion history, and hence $N$ and $\zeta$. For more on how these models are related see e.g.~\cite{Vernizzi:2003vs,Alabidi:2010ba,Elliston:2014zea}.

\subsection{Scale dependence of $\fnl$}\label{sec:scale-dep}

As defined in (\ref{zetaloc}), $\fnl$ is just a constant modulating the second order term in the expansion. Whilst this is a very good approximation in many models, it is generally not exact. Similarly to how the power spectrum usually has a spectral index of order the slow-roll parameters (reflecting evolution during inflation), the non-Gaussianity typically has a similar sized scale dependence. The analogy is not exact since there are simple models in which the scale invariance is exact (such as the quadratic curvaton model that we will study later), and there are other cases where the scale dependence is large, despite all fields obeying the slow-roll conditions.

In analogy to the spectral index, we define
\bea 
  \label{n_fnl_equil}
  \nfnl\equiv\frac{d\ln |\fnl(k)|}{d \ln k}.
\eea
Strictly speaking, this is only defined for an equilateral triangle, for which all three $k's$ are equal. However at lowest order, this spectral index is the same for any shape of triangle provided that one scales all three sides by the same ratio as shown in \cite{Byrnes:2010ft}. Corrections do become important in the case that one strongly deforms the triangle while changing its scaling \cite{Shandera:2010ei,Dias:2013rla}.

The most general formula for $\nfnl$ which also allows for a non-trivial field space metric is given in \cite{Byrnes:2012sc}. There it was shown that there are 3 effects which may give rise to scale dependence: 1) multiple field effects, from fields with different spectral indices contributing to the curvature perturbation 2) non-linearity of the field perturbation equation of motion, which is only absent in the case of a field with a quadratic potential and no backreaction from gravity and 3) a non-trivial field space metric. One example of a case with a scale independent $\fnl$ is the quadratic curvaton scenario, where the curvaton field's energy density is subdominant by construction and the perturbations from the inflaton field are neglected. However this is clearly an idealised case, and in general we should expect some scale dependence. 

How large can we expect the scale dependence to be? The answer is model dependent, but in some cases relatively simple results are known and may be used to gain intuition. In the case of a single-source inflation model (see the previous sub section), in which the corresponding field direction was an isocurvature mode during inflation (otherwise the non-Gaussianity is very small if the field is the inflaton)
 \bea
\label{fnlgnl_pivot}
\fnl&=&\frac56\frac{N_{\sigma\sigma}}{N_{\sigma}^2},
\\\label{ntnlcu} \nfnl 
&\simeq&
\frac{N_{\sigma}}{N_{\sigma\sigma}}\frac{V'''}{3H^2}\simeq
\frac56 \frac{{\rm sgn}(N_{\sigma})}{\fnl} \sqrt{\frac{r_{\rm T}}{8}}\frac{V'''}{3H^2}\ .  \label{nfnl:single}
\eea
In order for $\fnl$ to be observable, the term $1/\fnl$ must be less than unity. We also have the observational constraint from the tensor-to-scalar ratio that $\sqrt{r_{\rm T}/8}\lesssim0.1$. However the last term in (\ref{nfnl:single}) may be arbitrarily large, it is not a slow-roll parameter. Clearly it is zero in the case of a quadratic potential. The best studied case with non quadratic potentials, which often requires a numerical treatment, is the curvaton scenario. In this case, the scale-dependence is typically of a similar size to the slow-roll parameters but in some cases it may be much larger, even $\nfnl\sim1$ is possible, at which values the formalism used for the calculation becomes inaccurate. See Sec.~\ref{sec:selfinteractingcurvaton} for details.

If multiple-fields generate the curvature perturbation, e.g.~both the inflaton and the curvaton, then $\fnl$ will generically be scale dependent even when the non-Gaussian field has a quadratic potential. This is due to $\fnl$ being defined as the ratio of the bispectrum to the square of the power spectrum  (\ref{Bloc}). The bispectrum is only sourced by the non-Gaussian field, while the power spectrum is sourced by both fields. Unless the power spectra of both fields have the same scale-dependence, the relative importance of the two fields perturbations will depend on scale. On the scales where the non-Gaussian field is more important $\fnl$ is larger, while in the opposite case where the Gaussian (inflaton) field perturbations dominate $\fnl$ will be negligible. An explicit realisation of this case is presented in Sec.~\ref{sec:mixedinflatoncurvaton}.

\subsection{The trispectrum}\label{sec:trispectrum}

The (connected) four point function of the primordial curvature perturbation is defined by
\begin{equation}\label{Tdefn}
\la \zeta_{{\mathbf k_1}}\,\zeta_{{\mathbf k_2}}\, \zeta_{{\mathbf k_3}} \zeta_{{\mathbf
k_4}} \ra_c \equiv T_\zeta({\mathbf k_1},{\mathbf k_2},{\mathbf k_3}, {\mathbf k_4})
\picube \sdelta{{\mathbf k_1}+{\mathbf k_2}+{\mathbf k_3} +{\mathbf k_4}}\,,
\end{equation}
which using the $\delta N$ formalism, and assuming that the fields have a Gaussian distribution at Hubble exit see Sec.~\ref{sec:deltaN}, is given by
\begin{align}\label{tauNLgNLdefn} T_\zeta (\bkone,\bktwo,{\mathbf k_3},{\mathbf k_4}) =&
\tau_{NL}\left[P_\zeta(k_{13})P_\zeta(k_3)P_\zeta(k_4)+(11\,\,\rm{perms})\right] \nonumber \\
&+\frac{54}{25}g_{NL}\left[P_\zeta(k_2)P_\zeta(k_3)P_\zeta(k_4)+(3\,\,\rm{perms})\right]\,,
\end{align}
where $k_{13}=|{\mathbf k_1}+{\mathbf k_3}|$ and the trispectrum non-linearity parameters can be calculated using \cite{Alabidi:2005qi,Seery:2006js,Byrnes:2006vq}
\bea
\label{taunl}
\tau_{NL}&=&\frac{N_{ab}N_{ac}N_bN_c}{(N_d N_d)^3}\,, \\
\label{gnl}
g_{NL}&=&\frac{25}{54}\frac{N_{abc}N_a N_b N_c}{(N_d N_d)^3}\,.
\eea
Hence we see that the trispectrum depends on two non-linearity parameters (as opposed to one, $\fnl$, for the bispectrum), and they may be observationally distinguishable since they are prefactors of terms with different shape dependences in (\ref{tauNLgNLdefn}). The observational constraints on $\taunl$ are tighter than those on $\gnl$ because the pre factor  for $\taunl$ is large when either $k_i\rightarrow0$ or $k_{ij}\rightarrow0$, while it is large only in the former case for $\gnl$.

No constraint has yet been made with Planck data on $\gnl$, from WMAP9 data $\gnl=(-3.3\pm2.2)\times10^5$ \cite{Sekiguchi:2013hza}. From scale-dependent bias data, Giannantonio et al found $-5.6\times10^5 < \gnl < 5.1\times10^5$ (2-$\sigma$) assuming that $\fnl=0$, which weakens to $ -5.9\times10^5 < \gnl < 4.7\times10^5$ when marginalising over $\fnl$ \cite{Giannantonio:2013uqa}. Using the same technique, Leistedt et al recently found $-4.0 \times 10^5 < \gnl < 4.9\times 10^5$ (2-$\sigma$) when marginalising over $\fnl$ \cite{Leistedt:2014zqa}. The constraint on (positive definite) $\taunl$ comes from the Planck collaboration and is significantly tighter than the constraint on $\gnl$, being $\taunl<2800$ at $2-\sigma$ \cite{Ade:2013ydc}.

The derivation of (\ref{taunl}) and (\ref{gnl}) is similar to that of $\fnl$ in Sec.~\ref{sec:deltaN}, see \cite{Byrnes:2006vq} for details. The different forms of the contraction in the numerator follow due to $\taunl$ consisting of two first-order terms and two-second order terms, while $\gnl$ is made out of three first-order terms and one third-order term (corresponding to the third derivative of N in (\ref{gnl})).  In both cases  the total four-point function is sixth order in $\delta\phi$, or third order in the power spectrum. Thats why there is a pre factor proportional to the power spectrum cubed in from of the $\taunl$ and $\gnl$ terms in (\ref{tauNLgNLdefn}).

\subsection{Suyama-Yamaguchi inequality}

Applying the Cauchy-Schwarz inequality to (\ref{fnl}) and (\ref{taunl}) one may derive the Suyama -Yamaguchi inequality  \cite{Suyama:2007bg}
\bea\label{SYinequality} \tau_{\rm NL}\geq \left(\frac{6 f_{\rm NL}}{5}\right)^2. \eea
In the single-source limit the inequality saturates to an equality, and this is the most widely considered test of whether the curvature perturbation was generated by a single field (which we stress is not the same as asking whether inflation was driven by a single field, in fact the single field which generates the perturbations must not be the inflaton field in order for their to be any possibility of the trispectrum parameters being large enough to observe). 
Unfortunately the tight constraints on $\fnl$ and difficulty in constraining $\taunl$ means that we are only likely to be able to detect $\taunl$ if $\taunl\gg\fnl^2$, while testing the equality is beyond observational reach for the foreseeable future 
\cite{Biagetti:2012xy,Grassi:2013ana}.

\section{The curvaton scenario as a worked example}\label{sec:curvaton}

The curvaton scenario is arguably the most popular model for studying non-Gaussianity \cite{Enqvist:2001zp,Lyth:2001nq,Moroi:2001ct} (see also \cite{Mollerach:1989hu,Linde:1996gt} for earlier related work). The scenario is quite minimal in that it requires only the smallest possible extra complication for a model to be able to generate large local non-Gaussianity, and analytical solutions are possible, making this an excellent pedagogical example. We will consider this model in several parts, starting from the simplest case and then gradually dropping model assumptions and seeing how the picture becomes both richer and more complex. Often, dropping a model assumption also leads to the existence of a new observable, meaning that the different cases of the curvaton scenario are (at least in principle) distinguishable.

The curvaton is an additional scalar field present during inflation. In order to pick up scale invariant perturbations its mass must be light compared to the Hubble scale
, in addition it is required to have a subdominant energy density compared to the inflaton by assumption. Until section \ref{sec:selfinteractingcurvaton} we will assume it has a quadratic potential
\bea V=\frac12 m_\sigma^2\sigma^2. \eea
Due to the assumption that the curvatons energy density is small, its equation of motion for both the background and perturbation are the same during inflation, in the special case of a quadratic potential
\bea \ddot{\sigma}+3H\dot{\sigma}+V_{,\sigma}&=&0,  \label{curvaton:eom} \\ 
\ddot{\delta\sigma}+3H\dot{\delta\sigma}+V_{,\sigma\sigma}\delta\sigma&=&0. \label{curvaton:pertneom} \eea
Neglecting the kinetic energy density of the curvaton, its energy density perturbation is a constant, and is given by 
\bea \frac{\delta\rho_{\sigma}}{\rho_{\sigma}}\simeq\frac{V(\sigma+\delta\sigma)-V(\sigma)}{V(\sigma)}=2\frac{\delta\sigma}{\sigma}+ \left(\frac{\delta\sigma}{\sigma}\right)^2. \label{curv:deltarho}\eea
It is not possible to get the correct numerical factors using this simple treatment, but it is nontheless a useful approximation to relate the first term to the linear curvature perturbation caused by the curvaton, $\zeta^{(1)}_\sigma=\frac{\delta\sigma}{\sigma}$ and the second order term $\zeta^{(2)}_\sigma=\left(\frac{\delta\sigma}{\sigma}\right)^2$, so that (neglecting numerical factors), $\zeta_{\sigma}=\zeta^{(1)}_{\sigma}+\zeta^{(2)}_{\sigma}$ is a constant. However the total curvature perturbation is different and not conserved, it is proportional to $\Omega_{\sigma}=\rho_{\sigma}/\rho_{tot}$, assuming that the curvaton is the only perturbed component of the universe 
\bea \zeta\sim\Omega_{\sigma}\zeta_{\sigma}. \eea
We will later consider what changes when all of these assumptions are dropped. We may already learn two important lessons which remain true in more general contexts, $\fnl\propto\zeta^{(2)}/\zeta^{(1)2}\propto 1/\Omega_{\sigma}$, and $\gnl=0$. The first lesson is quite general, and states that even if the perturbations of a field as nearly Gaussian, as is the case for the curvaton, then an inefficient transfer of that fields curvature perturbation to the total curvature perturbation will generate non-Gaussianity. This follows since the transfer multiplies the curvature perturbation equally at all orders, making the second order term larger when compared to the square of the first order term. The second lesson is that quadratic potentials tend to only generate quadratic non-Gaussianity, notice that it is only for a quadratic potential that (\ref{curv:deltarho}) can be truncated at second order.

We now study the crucial parameter $\Omega_{\chi}$, which we have seen will affect both the amplitude of perturbations and the level of non-Gaussianity. During inflation, the curvaton will roll slowly due to the assumption that it is light, $\mchi\ll H$. However the total energy density is also slowly varying (due to the requirement that $\epsilon=-\dot{H}/H^2\ll1$), so $\Omega_\chi\simeq\rho_{\chi}/\rho_{\phi}\ll1$ is approximately constant (and very small) during this period. After inflation ends, the inflaton decays into radiation, which redshifts as $a^{-4}$. The curvaton will initially remain almost stationary, so that $\Omega_\chi\propto a^4$, until its mass become comparable to the Hubble rate at which time it will start to oscillate. Oscillations about a quadratic minimum correspond to a pressureless equation of state, so $\rho_\chi\propto a^{-3}$ and $\Omega_\chi\propto a$. During this period the relative energy density of the curvaton can grow a lot, potentially dominating the energy density of the universe if it either decays late enough or is much lighter than the Hubble parameter at the end of inflation. Finally the curvaton decays into radiation and thereafter $\Omega_\chi=\Omega_\chi|_{\rm decay}$ becomes a constant, and the perturbations in its field (which were initially isocurvature perturbations) have been converted into the primordial density perturbation. For a diagram displaying this process, see Fig.~4 of \cite{Dimopoulos:2010nq}.

Including numerical factors, see e.g.~\cite{Sasaki:2006kq} for a derivation, the curvature perturbation is given by
\bea\label{curvfNL} \zeta=\rdec\zeta^{(1)}_\sigma + \left(\frac{3}{2\rdec}-2-\rdec\right) \left(\zeta^{(1)}_\sigma\right)^2, \eea
where
\bea \rdec=\frac{3\rho_\sigma}{4\rho_{\rm radiation}+3\rho_{\sigma}}=\frac{3\Omega_\sigma}{4+3\Omega_\sigma}\Big|_{\rm decay}. \eea
From (\ref{curvfNL}) the full result for $\fnl$ is
\bea f_{NL}=\frac{5}{4 r_{\rm dec}}-\frac53-\frac56 r_{\rm dec}, \eea
which in the limit of a subdominant curvaton reduces to
\bea \fnl\propto \frac{1}{\rdec}\propto\frac{1}{\Omega_\sigma}, \qquad {\rm for}\;\; \rdec=\frac34\Omega_\sigma\ll1, \eea
in agreement with the arguments made above. In the opposite limit of a dominant curvaton at the decay time we find $\fnl=-5/4$ which is much less Gaussian than the slow-roll value predicted by single-field models (the Maldacena consistency relation, see (\ref{Maldacena})), but observationally indistinguishable from a Gaussian distribution. Observationally probing values of $\fnl\sim1$ is an important target for future experiments. Values of $\fnl$ even closer to zero are possible in the curvaton scenario, but require a finally tuned value of $\rdec$, while $\rdec=1$ is the asymptotic value in the limit that the curvaton decays late. 

If one had to choose one value of $\fnl$ as a prediction from the curvaton, it would be $\fnl=-5/4$. If the curvaton instead decayed when it was subdominant, then a large and positive value of $\fnl$ is predicted. The Planck constraint of $\fnl\lesssim10$ results in the bound $\rdec\gtrsim0.1$, notice that prior to an $\fnl$ constraint being made, $\rdec\sim10^{-5}$ was possible which would generate the correct power spectrum amplitude if $\delta\sigma/\sigma$ was approximately unity. Even without a detection of non-Gaussianity, the constraints have taught us under which circumstances the curvaton scenario can still be a viable model.



\subsection{Including the inflation field perturbations}\label{sec:mixedinflatoncurvaton}

So far we have assumed that the inflation field perturbations are negligible, however the field perturbations of any light scalar field are expected to have the same amplitude at horizon crossing, with the expectation value of their amplitude of perturbations at horizon crossing being
\bea \delta\phi=\delta\sigma=\frac{H}{2\pi}. \eea
The question is whether the curvature perturbation from the curvaton or the inflaton will be larger.

In this subsection, we will assume that the curvaton is subdominant at decay, $\rdec\ll1$ and we will neglect numerical factors of order unity. The curvature perturbation is given by
\bea \zeta\sim\zeta_{\phi} + \rdec (\zeta_\sigma+\zeta_\sigma^2),  \eea
where 
\bea \zeta_\phi\sim \frac{1}{\Mp\sqrt{\epsilon_*}}\delta\phi_*, \;\; \zeta_\sigma\sim\frac{\delta\sigma}{\sigma}\Big|_*, \eea
and quantities to be evaluated around horizon crossing are marked with a ``$*$".
The power spectra due to the two fields are
\bea P_{\zeta}^\phi\sim \frac{1}{\epsilon_*}\left(\frac{H_*}{2\pi}\right)^2, \qquad P_{\zeta}^{\sigma}\sim \Omega_{\sigma}^2 \frac{1}{\sigma_*^2}\left(\frac{H_*}{2\pi}\right)^2, \eea
and the total power spectrum is the sum, based on the good approximation that the perturbations of the two fields are uncorrelated \cite{Byrnes:2006fr}
\bea P_{\zeta}\equiv P_{\zeta}^{\rm total}=P_{\zeta}^{\phi}+P_{\zeta}^{\sigma}. \eea
Because the inflaton perturbations are Gaussian, the bispectrum is unchanged from the pure curvaton limit 
\bea B_\zeta=B_{\zeta}^{\sigma}\simeq\frac{1}{\rdec} \left(P_{\zeta}^{\sigma}\right)^2, \eea
but $\fnl$ is reduced because the power spectrum is enhanced by the Gaussian inflaton field perturbations
\bea \fnl=\frac56\frac{B_\zeta}{P_\zeta^2}= \frac56 \frac{B_{\zeta}^{\sigma}}{P_\zeta^2}\sim \frac{1}{\rdec}\left(\frac{P_{\zeta}^{\sigma}}{P_\zeta}\right)^2\propto k^{2(n_\sigma-n_s)}, \eea
where the spectral index $n_\sigma$ is defined by ${\cal P}_\zeta^{\sigma}\propto k^{n_\sigma-1}$. In the curvaton limit $n_\sigma\rightarrow n_s$ so $\fnl$ becomes scale-invariant (assuming that it has a quadratic potential). Alternatively in the limit that the curvaton power spectrum is scale invariant, we have 
\bea \nfnl \rightarrow -2(n_s-1)\simeq 0.1. \eea

The tensor-to-scalar ratio is also reduced compared to the single-field inflation limit, whose value is $r_T=16\epsilon_*$, to
\bea r_T=16\epsilon_* \frac{P_{\zeta}^{\phi}}{P_\zeta}.  \eea
In the curvaton limit $P_{\zeta}^{\phi}\ll P_\zeta$, so $r_T$ is expected to be negligible (remember that $\epsilon<1$ is required in order to have inflation). 

The trispectrum parameter $\taunl$ is instead enhanced compared to $\fnl^2$, we have
\bea  \taunl=\frac{P_\zeta}{P^\sigma_\zeta}\left(\frac{6\fnl}{5}\right)^2 > \left(\frac{6\fnl}{5}\right)^2, \eea
which obeys the Suyama-Yamaguchi inequality (\ref{SYinequality}). Notice that $\taunl$ is also reduced by the addition of Gaussian inflaton perturbations, but by a lesser amount than $\fnl^2$. Finally $\gnl$ will remain negligible under the additional contribution of Gaussian perturbations from the inflation field.

\subsection{The self-interacting curvaton}\label{sec:selfinteractingcurvaton}

Due to our assumption of a quadratic potential, the density and curvature perturbation are only present up to second order. There are subleading corrections in $1/\rdec$ which give rise to higher order terms, but they are too small to ever be observed.  For non-quadratic potentials there is a potentially large third order term in (\ref{curv:deltarho}), which can be constrained or measured by observations of $\gnl$. In general we may expect $\gnl\sim\taunl\sim\fnl^2$ from the curvaton scenario, unless it has a quadratic potential.  

There is no fundamental reason why a quadratic potential is more ``natural'' or likely than non-quadratic cases, but most of the literature has focused on quadratic potentials due to the great computational simplicity which follows and allows analytical results to be found. When we include self interactions of the curvaton, i.e.~extensions beyond  a quadratic potential, the equation of motion for the curvaton becomes non-linear (see (\ref{curvaton:eom}) and (\ref{curvaton:pertneom})) and $\delta\sigma/\sigma$ is no longer conserved. This can lead to very interesting behaviour \cite{Enqvist:2009ww}, such as strongly scale-dependent non-Gaussianity. For certain (finely tuned) initial conditions one has $\nfnl\gg {\cal O}(\epsilon, \eta)$, turning this into an additional potential observable \cite{Byrnes:2011gh}. Another difference from the quadratic case is that there is no lower bound of $\fnl$, and the value of $\fnl$ in the limit $\rdec\rightarrow1$ depends on the curvaton's potential. Generically the prediction does remain that $\fnl$ is of order unity, but not exactly $-5/4$.

\subsection{Curvaton scenario summary}

We close this section with a table demonstrating how the many different realisations of the curvaton scenario could be distinguished using multiple non-Gaussian observables. This has provided us with a concrete realisation of how powerful an observation of non-Gaussianity would be, as well as how even the current constraints without a detection provide interesting information about the curvaton. 

\begin{table}[h]
\begin{tabular}{|l|l|l|l|l|l|}
\hline
Scenario                                                                                                         & Curvaton potential & $\fnl$         & $\nfnl$                & $\taunl/(6\fnl/5)^2$ & $\gnl$               \\ \hline
\multirow{2}{*}{\begin{tabular}[c]{@{}l@{}}pure curvaton \\ scenario \end{tabular}} & quadratic          & $\geq-\frac54$ & 0                      & 1                    & small \\ \cline{2-6} 
                                                                                                                 & non-quadratic      & unrestricted   & potentially large      & 1                    & potentially large  \\ \hline
\multirow{2}{*}{\begin{tabular}[c]{@{}l@{}}mixed inflaton-\\ curvaton\end{tabular}}                              & quadratic          & $\geq-\frac54$ & slow-roll$\sim10^{-2}$ & $\geq1$              &  small \\ \cline{2-6} 
                                                                                                                 & non-quadratic      & unrestricted   & potentially large      & $\geq1$              & potentially large  \\ \hline
\end{tabular}
\caption{Four different curvaton scenarios, with a quadratic or non-quadratic curvaton potential and either including or neglecting the inflaton perturbations. The three columns of non-Gaussian observables show how we can (in principle) distinguish between the different cases.}
\label{table:curvaton}
\end{table}

\section{Frequently Asked Questions}\label{sec:faqs}

Note that these lectures were written after the first Planck cosmology data release in Spring 2013, but before the final data release which is expected in late 2014 or 2015.

\begin{itemize}

\item {\bf Do the Planck non-Gaussianity constraints imply that there is negligible non-Gaussianity?}

Not really. For the local model of non-Gaussianity, they do imply the sky is over 99.9\% Gaussian, which is a remarkable result. For other templates, the constraint could be much weaker. But the constraint $|\fnlloc|\lesssim10$ are still two to three orders of magnitude larger than the single-field consistency relation for the squeezed limit of the bispectrum, $\fnl\simeq n_s-1$. Clearly a large window is left for models which strongly deviate from this consistency relation, but have a level of non-Gaussianity which is not yet detectable.

\item {\bf Do the Planck non-Gaussianity constraints imply that alternatives to single field inflation are strongly disfavoured?}

No. Single field inflation remains consistent with the observations, which does suggest they should be preferred from a Bayesian/Occams razor perspective. This was also true before we had Planck results. However it is important to bear two points in mind: 1) A model which is parametrised with the fewest parameters might not be the simplest or most natural from a model building perspective, (we know little about physics at the inflationary energy scale) and 2) there are many multiple field models which predict non-Gaussianity with $|\fnl|\ll1$, and hence are far from being ruled out.

\item {\bf Is there a natural target for future non-Gaussianity experiments?}

Yes, to a certain extent. Several models which convert an isocurvature perturbation present during inflation into the primordial adiabatic perturbation after inflation have a large parameter range in which $|\fnlloc|\sim1$. For example, the simplest version of the curvaton scenario, in which the curvaton potential is quadratic and it is dominant at the decay time (which it will be the case if it decays sufficiently late) makes a definite prediction, $\fnlloc=-5/4$. Similarly, a particularly simple realisation of modulated reheating predicts $\fnlloc=5/2$. Hence having an experiment which is capable of discriminating between $\fnlloc=1$ and $\fnlloc=0$ would have great value in disfavouring popular non-Gaussian models. See the next question for an idea of when we might reach this target. 

For the equilateral model of non-Gaussianity, $\fnleq\sim1$ is also a natural target for testing models with a non-canonical kinetic term \cite{Baumann:2014cja}.

\item {\bf What are the prospects for future non-Gaussianity measurements?}

The final Planck data release, which will contain double the observation time compared to the first release as well as Planck polarisation data, is expected to only lead to a relatively modest improvement to the $\fnl$ constraints, about $20\%$, compared to a factor of two for several other cosmological parameters including the spectral index. The next significant improvement in the constraint for $\fnlloc$ is expected around 2020 from the Euclid survey, which is forecasted to reach an error bar of around 2-3 (i.e.~around a factor of two tighter than Planck) \cite{Giannantonio:2011ya}. Combined with SKA data in perhaps a decade, the constraints are forecasted to be able to distinguish $\fnlloc=0$ from $\fnlloc=\pm1$ at about $3-\sigma$ confidence \cite{Yamauchi:2014ioa}. Beyond this, there is no clear timeline to future experiments which will have even tighter constraints, although several experiments have been proposed, for example Core, Prism and Pixie which would measure the CMB to greater accuracy and to smaller scales.

\item {\bf Which forms of non-Gaussianity can we best constrain with future experiments?}

Currently, the only concrete expectation for a significant improvement in non-Gaussianity constraints comes from the Euclid satellite. The forecasts have mainly been made for the scale dependent halo bias, which is sensitive to the squeezed limit of the bispectrum and hence primarily to local non-Gaussianity. 
The prospects for the other shapes is weaker, but limited work has been done on studying the galaxy bispectrum and using this as an estimator which could potentially improve sensitivity to all shapes of the bispectrum. This work is very challenging since the secondary signal from non-linear collapse is much larger than the primordial signal (implying observations will have to deal with many potentially large systematic effects). Even starting with Gaussian initial conditions, structure formation is a challenging topic.

\end{itemize}

\section{Conclusions and future outlook}\label{sec:conclusions}

Non-Gaussianity is a window on to the physics of the very early universe. The distribution of the primordial perturbations contains much more information if it is non-Gaussian, providing signatures on to the high energy physics of inflation. 

We have provided an introduction to the field of primordial non-Gaussianity. In contrast to a Gaussian perturbation, which is simple to describe and has only a variance as a free parameter, a non-Gaussian perturbation could be anything else and have any number of free parameters. Fortunately well motivated models of the early universe tend to predict a reasonably small number of non-Gaussian templates, which may be efficiently parametrised in terms of the three-point function of the curvature perturbation (or temperature perturbation on the CMB). For  a Gaussian perturbation all information is included in the two-point function (i.e.~the power spectrum) and the three-point function is zero. Hence any detection of the three point function would prove the primordial perturbations were non-Gaussian and the cosmology community has made a great effort to studying the bispectrum. This effort includes calculating the amplitude and shape dependence for classes of inflationary models as well as constraining these templates against observations. Planck has made the tightest ever constraints of the bispectrum, which are significantly tighter than anything which existed before. Although there was no detection of non-Gaussianity, the constraints already rule out or put some inflationary models under observational pressure, and constrain the primordial perturbations to be more than $99.9\%$ Gaussian. 

The simplest models of single-field slow-roll inflation predict a much smaller deviation from Gaussianity, to a level which is probably too small to ever be tested. However for other models of inflation there is still hope to for a future detection of non-Gaussianity, for example we have seen how $\fnl=-5/4$ is a natural prediction of the curvaton scenario, and this value is within an order of magnitude of the current constraints. Improving the constraints so much will not be possible with CMB data, but may be possible with the next generation of large scale structure surveys in about a decade. They probe three dimensional information which allows them to alleviate the cosmic variance limits which the last scattering surface of the CMB suffers from, since we only have one sky to observe. However large scale structure has undergone more evolution through gravitational collapse than the CMB, and the non-linear equations of GR have made the perturbations less linear. The hard task is then to separate the primordial non-Gaussianity from the already detected secondary non-Gaussianities. 

A detection of primordial non-Gaussianity would either be a signature of non-linear physics acting during the generation of the primordial perturbations or a non-linear transformation of the primordial perturbation between the initial scalar field perturbation and the curvature perturbation. An example of the former case is inflation with a reduced sound speed of perturbations due to a non-canonical kinetic term, while an in depth study of the latter case was made for the curvaton scenario. A general formalism ($\delta N$) was also provided for calculating this non-linear transformation, which for many models allows a study of the perturbations even to non-linear order just by calculating background quantities. This remarkable simplification is possible for the (many) models in which the perturbations are Gaussian at Hubble crossing during inflation and leads to local non-Gaussianity, described by a combination of a Gaussian perturbation and chi-squared non-Gaussianity.

While nature has not provided many clues to the physics of inflation, the search is continuing in a large way. The primordial perturbations have been convincingly demonstrated to deviate from scale invariance at a level consistent with slow-roll inflation. The search for primordial tensor perturbation, non-Gaussianity, isocurvature perturbations and features in the power spectrum, etc, continues. Whether or not these extra signatures are detected, it is only by studying a large range of inflationary models that we learn what to search for with the ever improving data sets, and what the constraints tell us about the first fraction of a second after the big bang.

\acknowledgements{CB is supported by a Royal Society University Research Fellowship. I wish to thank the organisers of the second JPBCosmo school for the invitation to hold this lecture course in the shadow of the beautiful Pedra Azul mountain in Espirito Santo, Brasil. Special thanks are due to Hermano Endlich Schneider Velten and his family for hospitality before and after the school.}

\bibliographystyle{unsrt}
\bibliography{review-brasil.bib}

\begin{thebibliography}{10}

\bibitem{Komatsu:2009kd}
E.~Komatsu, N.~Afshordi, N.~Bartolo, D.~Baumann, J.R. Bond, et~al.
\newblock {Non-Gaussianity as a Probe of the Physics of the Primordial Universe
  and the Astrophysics of the Low Redshift Universe}.
\newblock 2009.

\bibitem{Maldacena:2002vr}
Juan~Martin Maldacena.
\newblock {Non-Gaussian features of primordial fluctuations in single field
  inflationary models}.
\newblock {\em JHEP}, 0305:013, 2003.

\bibitem{Komatsu:2010hc}
Eiichiro Komatsu.
\newblock {Hunting for Primordial Non-Gaussianity in the Cosmic Microwave
  Background}.
\newblock {\em Class.Quant.Grav.}, 27:124010, 2010.

\bibitem{Yadav:2010fz}
Amit~P.S. Yadav and Benjamin~D. Wandelt.
\newblock {Primordial Non-Gaussianity in the Cosmic Microwave Background}.
\newblock {\em Adv.Astron.}, 2010:565248, 2010.

\bibitem{Liguori:2010hx}
Michele Liguori, Emiliano Sefusatti, James~R. Fergusson, and E.P.S. Shellard.
\newblock {Primordial non-Gaussianity and Bispectrum Measurements in the Cosmic
  Microwave Background and Large-Scale Structure}.
\newblock {\em Adv.Astron.}, 2010:980523, 2010.

\bibitem{Desjacques:2010nn}
Vincent Desjacques and Uros Seljak.
\newblock {Primordial non-Gaussianity in the large scale structure of the
  Universe}.
\newblock {\em Adv.Astron.}, 2010:908640, 2010.

\bibitem{Verde:2010wp}
Licia Verde.
\newblock {Non-Gaussianity from Large-Scale Structure Surveys}.
\newblock {\em Adv.Astron.}, 2010:768675, 2010.

\bibitem{Bartolo:2010qu}
N.~Bartolo, S.~Matarrese, and A.~Riotto.
\newblock {Non-Gaussianity and the Cosmic Microwave Background Anisotropies}.
\newblock {\em Adv.Astron.}, 2010:157079, 2010.

\bibitem{Abramo:2010gk}
L.~Raul Abramo and Thiago~S. Pereira.
\newblock {Testing gaussianity, homogeneity and isotropy with the cosmic
  microwave background}.
\newblock {\em Adv.Astron.}, 2010:378203, 2010.

\bibitem{Copi:2010na}
Craig~J. Copi, Dragan Huterer, Dominik~J. Schwarz, and Glenn~D. Starkman.
\newblock {Large angle anomalies in the CMB}.
\newblock {\em Adv.Astron.}, 2010:847541, 2010.

\bibitem{Chen:2010xka}
Xingang Chen.
\newblock {Primordial Non-Gaussianities from Inflation Models}.
\newblock {\em Adv.Astron.}, 2010:638979, 2010.

\bibitem{Wands:2010af}
David Wands.
\newblock {Local non-Gaussianity from inflation}.
\newblock {\em Class.Quant.Grav.}, 27:124002, 2010.

\bibitem{Byrnes:2010em}
Christian~T. Byrnes and Ki-Young Choi.
\newblock {Review of local non-Gaussianity from multi-field inflation}.
\newblock {\em Adv.Astron.}, 2010:724525, 2010.

\bibitem{Ade:2013zuv}
P.A.R. Ade et~al.
\newblock {Planck 2013 results. XVI. Cosmological parameters}.
\newblock {\em Astron.Astrophys.}, 571:A16, 2014.

\bibitem{Bennett:2010jb}
C.L. Bennett, R.S. Hill, G.~Hinshaw, D.~Larson, K.M. Smith, et~al.
\newblock {Seven-Year Wilkinson Microwave Anisotropy Probe (WMAP) Observations:
  Are There Cosmic Microwave Background Anomalies?}
\newblock {\em Astrophys.J.Suppl.}, 192:17, 2011.

\bibitem{Komatsu:2001rj}
Eiichiro Komatsu and David~N. Spergel.
\newblock {Acoustic signatures in the primary microwave background bispectrum}.
\newblock {\em Phys.Rev.}, D63:063002, 2001.

\bibitem{Senatore:2009gt}
Leonardo Senatore, Kendrick~M. Smith, and Matias Zaldarriaga.
\newblock {Non-Gaussianities in Single Field Inflation and their Optimal Limits
  from the WMAP 5-year Data}.
\newblock {\em JCAP}, 1001:028, 2010.

\bibitem{Chen:2005fe}
Xingang Chen.
\newblock {Running non-Gaussianities in DBI inflation}.
\newblock {\em Phys.Rev.}, D72:123518, 2005.

\bibitem{ArmendarizPicon:1999rj}
C.~Armendariz-Picon, T.~Damour, and Viatcheslav~F. Mukhanov.
\newblock {k - inflation}.
\newblock {\em Phys.Lett.}, B458:209--218, 1999.

\bibitem{Silverstein:2003hf}
Eva Silverstein and David Tong.
\newblock {Scalar speed limits and cosmology: Acceleration from D-cceleration}.
\newblock {\em Phys.Rev.}, D70:103505, 2004.

\bibitem{Alishahiha:2004eh}
Mohsen Alishahiha, Eva Silverstein, and David Tong.
\newblock {DBI in the sky}.
\newblock {\em Phys.Rev.}, D70:123505, 2004.

\bibitem{Baumann:2014nda}
Daniel Baumann and Liam McAllister.
\newblock {Inflation and String Theory}.
\newblock 2014.

\bibitem{Lidsey:2007gq}
James~E. Lidsey and Ian Huston.
\newblock {Gravitational wave constraints on Dirac-Born-Infeld inflation}.
\newblock {\em JCAP}, 0707:002, 2007.

\bibitem{Christopherson:2008ry}
Adam~J. Christopherson and Karim~A. Malik.
\newblock {The non-adiabatic pressure in general scalar field systems}.
\newblock {\em Phys.Lett.}, B675:159--163, 2009.

\bibitem{Ade:2013ydc}
P.A.R. Ade et~al.
\newblock {Planck 2013 Results. XXIV. Constraints on primordial
  non-Gaussianity}.
\newblock 2013.

\bibitem{Chen:2006xjb}
Xingang Chen, Richard Easther, and Eugene~A. Lim.
\newblock {Large Non-Gaussianities in Single Field Inflation}.
\newblock {\em JCAP}, 0706:023, 2007.

\bibitem{Chen:2006nt}
Xingang Chen, Min-xin Huang, Shamit Kachru, and Gary Shiu.
\newblock {Observational signatures and non-Gaussianities of general single
  field inflation}.
\newblock {\em JCAP}, 0701:002, 2007.

\bibitem{Holman:2007na}
R.~Holman and Andrew~J. Tolley.
\newblock {Enhanced Non-Gaussianity from Excited Initial States}.
\newblock {\em JCAP}, 0805:001, 2008.

\bibitem{Meerburg:2009ys}
Pieter~Daniel Meerburg, Jan~Pieter van~der Schaar, and Pier~Stefano Corasaniti.
\newblock {Signatures of Initial State Modifications on Bispectrum Statistics}.
\newblock {\em JCAP}, 0905:018, 2009.

\bibitem{Ringeval:2010ca}
Christophe Ringeval.
\newblock {Cosmic strings and their induced non-Gaussianities in the cosmic
  microwave background}.
\newblock {\em Adv.Astron.}, 2010:380507, 2010.

\bibitem{Ade:2013xla}
P.A.R. Ade et~al.
\newblock {Planck 2013 results. XXV. Searches for cosmic strings and other
  topological defects}.
\newblock 2013.

\bibitem{2012PhRvL.108w1301T}
P.~{Trivedi}, T.~R. {Seshadri}, and K.~{Subramanian}.
\newblock {Cosmic Microwave Background Trispectrum and Primordial Magnetic
  Field Limits}.
\newblock {\em Physical Review Letters}, 108(23):231301, June 2012.

\bibitem{Durrer:2013pga}
Ruth Durrer and Andrii Neronov.
\newblock {Cosmological Magnetic Fields: Their Generation, Evolution and
  Observation}.
\newblock {\em Astron.Astrophys.Rev.}, 21:62, 2013.

\bibitem{Fergusson:2008ra}
J.R. Fergusson and E.P.S. Shellard.
\newblock {The shape of primordial non-Gaussianity and the CMB bispectrum}.
\newblock {\em Phys.Rev.}, D80:043510, 2009.

\bibitem{Starobinsky:1986fxa}
Alexei~A. Starobinsky.
\newblock {Multicomponent de Sitter (Inflationary) Stages and the Generation of
  Perturbations}.
\newblock {\em JETP Lett.}, 42:152--155, 1985.

\bibitem{Sasaki:1995aw}
Misao Sasaki and Ewan~D. Stewart.
\newblock {A General analytic formula for the spectral index of the density
  perturbations produced during inflation}.
\newblock {\em Prog.Theor.Phys.}, 95:71--78, 1996.

\bibitem{Sasaki:1998ug}
Misao Sasaki and Takahiro Tanaka.
\newblock {Superhorizon scale dynamics of multiscalar inflation}.
\newblock {\em Prog.Theor.Phys.}, 99:763--782, 1998.

\bibitem{Lyth:2004gb}
David~H. Lyth, Karim~A. Malik, and Misao Sasaki.
\newblock {A General proof of the conservation of the curvature perturbation}.
\newblock {\em JCAP}, 0505:004, 2005.

\bibitem{Wands:2000dp}
David Wands, Karim~A. Malik, David~H. Lyth, and Andrew~R. Liddle.
\newblock {A New approach to the evolution of cosmological perturbations on
  large scales}.
\newblock {\em Phys.Rev.}, D62:043527, 2000.

\bibitem{Byrnes:2006fr}
Christian~T. Byrnes and David Wands.
\newblock {Curvature and isocurvature perturbations from two-field inflation in
  a slow-roll expansion}.
\newblock {\em Phys.Rev.}, D74:043529, 2006.

\bibitem{Lyth:2005fi}
David~H. Lyth and Yeinzon Rodriguez.
\newblock {The Inflationary prediction for primordial non-Gaussianity}.
\newblock {\em Phys.Rev.Lett.}, 95:121302, 2005.

\bibitem{Creminelli:2004yq}
Paolo Creminelli and Matias Zaldarriaga.
\newblock {Single field consistency relation for the 3-point function}.
\newblock {\em JCAP}, 0410:006, 2004.

\bibitem{Dvali:2003em}
Gia Dvali, Andrei Gruzinov, and Matias Zaldarriaga.
\newblock {A new mechanism for generating density perturbations from
  inflation}.
\newblock {\em Phys.Rev.}, D69:023505, 2004.

\bibitem{Zaldarriaga:2003my}
Matias Zaldarriaga.
\newblock {Non-Gaussianities in models with a varying inflaton decay rate}.
\newblock {\em Phys.Rev.}, D69:043508, 2004.

\bibitem{Suyama:2007bg}
Teruaki Suyama and Masahide Yamaguchi.
\newblock {Non-Gaussianity in the modulated reheating scenario}.
\newblock {\em Phys.Rev.}, D77:023505, 2008.

\bibitem{Ichikawa:2008ne}
Kazuhide Ichikawa, Teruaki Suyama, Tomo Takahashi, and Masahide Yamaguchi.
\newblock {Primordial Curvature Fluctuation and Its Non-Gaussianity in Models
  with Modulated Reheating}.
\newblock {\em Phys.Rev.}, D78:063545, 2008.

\bibitem{Bernardeau:2002jy}
Francis Bernardeau and Jean-Philippe Uzan.
\newblock {NonGaussianity in multifield inflation}.
\newblock {\em Phys.Rev.}, D66:103506, 2002.

\bibitem{Bernardeau:2002jf}
Francis Bernardeau and Jean-Philippe Uzan.
\newblock {Inflationary models inducing non-Gaussian metric fluctuations}.
\newblock {\em Phys.Rev.}, D67:121301, 2003.

\bibitem{Lyth:2005qk}
David~H. Lyth.
\newblock {Generating the curvature perturbation at the end of inflation}.
\newblock {\em JCAP}, 0511:006, 2005.

\bibitem{Huang:2009vk}
Qing-Guo Huang.
\newblock {A Geometric description of the non-Gaussianity generated at the end
  of multi-field inflation}.
\newblock {\em JCAP}, 0906:035, 2009.

\bibitem{Alabidi:2006hg}
Laila Alabidi.
\newblock {Non-gaussianity for a Two Component Hybrid Model of Inflation}.
\newblock {\em JCAP}, 0610:015, 2006.

\bibitem{Byrnes:2008wi}
Christian~T. Byrnes, Ki-Young Choi, and Lisa~M.H. Hall.
\newblock {Conditions for large non-Gaussianity in two-field slow-roll
  inflation}.
\newblock {\em JCAP}, 0810:008, 2008.

\bibitem{Peterson:2010mv}
Courtney~M. Peterson and Max Tegmark.
\newblock {Non-Gaussianity in Two-Field Inflation}.
\newblock {\em Phys.Rev.}, D84:023520, 2011.

\bibitem{Wang:2010si}
Tower Wang.
\newblock {Note on Non-Gaussianities in Two-field Inflation}.
\newblock {\em Phys.Rev.}, D82:123515, 2010.

\bibitem{Elliston:2011dr}
Joseph Elliston, David~J. Mulryne, David Seery, and Reza Tavakol.
\newblock {Evolution of fNL to the adiabatic limit}.
\newblock {\em JCAP}, 1111:005, 2011.

\bibitem{Vernizzi:2003vs}
Filippo Vernizzi.
\newblock {Cosmological perturbations from varying masses and couplings}.
\newblock {\em Phys.Rev.}, D69:083526, 2004.

\bibitem{Alabidi:2010ba}
Laila Alabidi, Karim Malik, Christian~T. Byrnes, and Ki-Young Choi.
\newblock {How the curvaton scenario, modulated reheating and an inhomogeneous
  end of inflation are related}.
\newblock {\em JCAP}, 1011:037, 2010.

\bibitem{Elliston:2014zea}
Joseph Elliston, Stefano Orani, and David~J. Mulryne.
\newblock {General analytic predictions of two-field inflation and perturbative
  reheating}.
\newblock {\em Phys.Rev.}, D89:103532, 2014.

\bibitem{Byrnes:2010ft}
Christian~T. Byrnes, Mischa Gerstenlauer, Sami Nurmi, Gianmassimo Tasinato, and
  David Wands.
\newblock {Scale-dependent non-Gaussianity probes inflationary physics}.
\newblock {\em JCAP}, 1010:004, 2010.

\bibitem{Shandera:2010ei}
Sarah Shandera, Neal Dalal, and Dragan Huterer.
\newblock {A generalized local ansatz and its effect on halo bias}.
\newblock {\em JCAP}, 1103:017, 2011.

\bibitem{Dias:2013rla}
Mafalda Dias, Raquel~H. Ribeiro, and David Seery.
\newblock {Scale-dependent bias from multiple-field inflation}.
\newblock {\em Phys.Rev.}, D87:107301, 2013.

\bibitem{Byrnes:2012sc}
Christian~T. Byrnes and Jinn-Ouk Gong.
\newblock {General formula for the running of fNL}.
\newblock {\em Phys.Lett.}, B718:718--721, 2013.

\bibitem{Alabidi:2005qi}
Laila Alabidi and David~H. Lyth.
\newblock {Inflation models and observation}.
\newblock {\em JCAP}, 0605:016, 2006.

\bibitem{Seery:2006js}
David Seery and James~E. Lidsey.
\newblock {Non-Gaussianity from the inflationary trispectrum}.
\newblock {\em JCAP}, 0701:008, 2007.

\bibitem{Byrnes:2006vq}
Christian~T. Byrnes, Misao Sasaki, and David Wands.
\newblock {The primordial trispectrum from inflation}.
\newblock {\em Phys.Rev.}, D74:123519, 2006.

\bibitem{Sekiguchi:2013hza}
Toyokazu Sekiguchi and Naoshi Sugiyama.
\newblock {Optimal constraint on $g_{NL}$ from CMB}.
\newblock {\em JCAP}, 1309:002, 2013.

\bibitem{Giannantonio:2013uqa}
Tommaso Giannantonio, Ashley~J. Ross, Will~J. Percival, Robert Crittenden,
  David Bacher, et~al.
\newblock {Improved Primordial Non-Gaussianity Constraints from Measurements of
  Galaxy Clustering and the Integrated Sachs-Wolfe Effect}.
\newblock {\em Phys.Rev.}, D89:023511, 2014.

\bibitem{Leistedt:2014zqa}
Boris Leistedt, Hiranya~V. Peiris, and Nina Roth.
\newblock {Constraints on primordial non-Gaussianity from 800,000 photometric
  quasars}.
\newblock 2014.

\bibitem{Biagetti:2012xy}
Matteo Biagetti, Vincent Desjacques, and Antonio Riotto.
\newblock {Testing Multi-Field Inflation with Galaxy Bias}.
\newblock 2012.

\bibitem{Grassi:2013ana}
Alessandra Grassi, Lavinia Heisenberg, Chris~T. Byrnes, and Bjoern~Malte
  Schaefer.
\newblock {A test of the Suyama-Yamaguchi inequality from weak lensing}.
\newblock 2013.

\bibitem{Enqvist:2001zp}
Kari Enqvist and Martin~S. Sloth.
\newblock {Adiabatic CMB perturbations in pre - big bang string cosmology}.
\newblock {\em Nucl.Phys.}, B626:395--409, 2002.

\bibitem{Lyth:2001nq}
David~H. Lyth and David Wands.
\newblock {Generating the curvature perturbation without an inflaton}.
\newblock {\em Phys.Lett.}, B524:5--14, 2002.

\bibitem{Moroi:2001ct}
Takeo Moroi and Tomo Takahashi.
\newblock {Effects of cosmological moduli fields on cosmic microwave
  background}.
\newblock {\em Phys.Lett.}, B522:215--221, 2001.

\bibitem{Mollerach:1989hu}
Silvia Mollerach.
\newblock {Isocurvature Baryon Perturbations and Inflation}.
\newblock {\em Phys.Rev.}, D42:313--325, 1990.

\bibitem{Linde:1996gt}
Andrei~D. Linde and Viatcheslav~F. Mukhanov.
\newblock {Nongaussian isocurvature perturbations from inflation}.
\newblock {\em Phys.Rev.}, D56:535--539, 1997.

\bibitem{Dimopoulos:2010nq}
Konstantinos Dimopoulos.
\newblock {The quantum origin of cosmic structure: theory and observations}.
\newblock {\em J.Phys.Conf.Ser.}, 283:012010, 2011.

\bibitem{Sasaki:2006kq}
Misao Sasaki, Jussi Valiviita, and David Wands.
\newblock {Non-Gaussianity of the primordial perturbation in the curvaton
  model}.
\newblock {\em Phys.Rev.}, D74:103003, 2006.

\bibitem{Enqvist:2009ww}
Kari Enqvist, Sami Nurmi, Olli Taanila, and Tomo Takahashi.
\newblock {Non-Gaussian Fingerprints of Self-Interacting Curvaton}.
\newblock {\em JCAP}, 1004:009, 2010.

\bibitem{Byrnes:2011gh}
Christian~T. Byrnes, Kari Enqvist, Sami Nurmi, and Tomo Takahashi.
\newblock {Strongly scale-dependent polyspectra from curvaton
  self-interactions}.
\newblock {\em JCAP}, 1111:011, 2011.

\bibitem{Baumann:2014cja}
Daniel Baumann, Daniel Green, and Rafael~A. Porto.
\newblock {B-modes and the Nature of Inflation}.
\newblock 2014.

\bibitem{Giannantonio:2011ya}
Tommaso Giannantonio, Cristiano Porciani, Julien Carron, Adam Amara, and
  Annalisa Pillepich.
\newblock {Constraining primordial non-Gaussianity with future galaxy surveys}.
\newblock {\em Mon.Not.Roy.Astron.Soc.}, 422:2854--2877, 2012.

\bibitem{Yamauchi:2014ioa}
Daisuke Yamauchi, Keitaro Takahashi, and Masamune Oguri.
\newblock {Constraining primordial non-Gaussianity via multi-tracer technique
  with Euclid and SKA}.
\newblock 2014.

\end{thebibliography}



\end{document}